# Focussing Light Using Negative Refraction

*JB Pendry and SA Ramakrishna*
*The Blackett Laboratory*
*Imperial College London, SW7 2AZ*
*UK*

*Abstract*

A slab of negatively refracting material, thickness $d$, can focus an image at a distance $2d$ from the object. The negative slab cancels an equal thickness of positive space. This result is a special case of a much wider class of focussing: any medium can be optically cancelled by an equal thickness of material constructed to be an inverted mirror image of the medium, with $\varepsilon, \mu$ reversed in sign. We introduce the powerful technique of coordinate transformation, mapping a known system into an equivalent system, to extend the result to a much wider class of structures including cylinders, spheres, and intersecting planes and hence show how to produce magnified images. All the images are 'perfect' in the sense that both the near and far fields are brought to a focus and hence reveal sub wavelength details.

## 1. Introduction

The transfer of an image from one location to another is conventionally done with lenses. An electromagnetic signal defined in one plane is refocused by a lens system on another plane or surface. This well developed technology suffers from some fundamental limitations of resolution. No details finer than the wavelength of radiation can be resolved. Some time ago Veselago [1] observed that if it were possible to realise materials with the property,

$$\varepsilon = \mu = -1 \tag{1}$$

then a refractive index of $n = -1$ would result and a slab of the material would focus light. For example an object placed a distance $d$ in front of a slab thickness $d$ would be refocused on the far side of the slab: see figure 1.

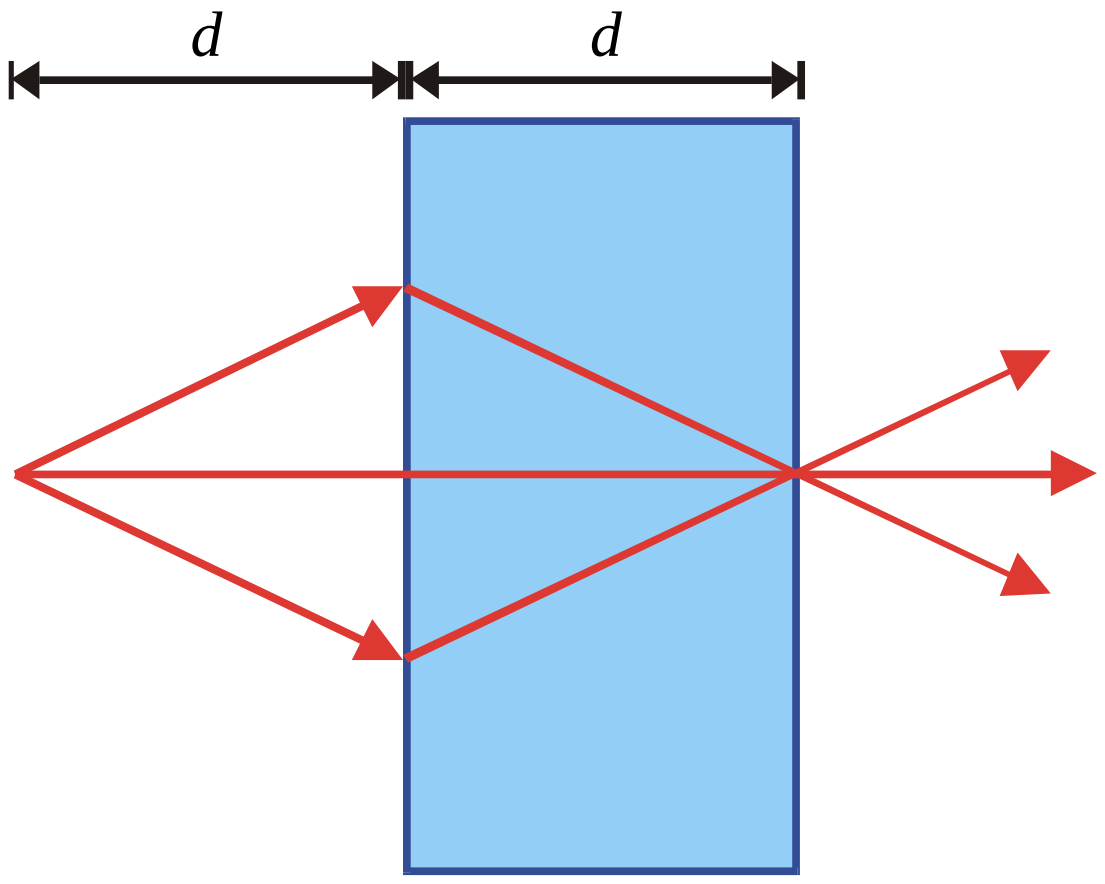

**Figure 1.** A slab of $n = -1$ of thickness $d$ focuses over a distance $2d$, as if the slab 'cancels out' or compensates for an equal thickness of the vacuum. The slab and the vacuum are complementary to each other.

The properties of a conventional lens are described by the focal length and by changing the position of an object relative to the lens, images of various magnifications can be obtained. The negative refraction lens works according to quite different principles. Here we build on the concept that a negatively refracting slab, $n = -1$, is in some sense complementary to an equal thickness of vacuum and cancels its presence. Optically speaking it is as if we were to remove a slab of space thickness $2d$. This cancellation is unusually complete because the compensating effect of the slab extends not only to the radiative component of the field, but also to the evanescent near field which conveys the sub-wavelength details of the image. Ordinary lenses do not capture the near field hence the limitations to their resolution. This extraordinary property of the Veselago lens was pointed out in 2000 [2]. Lenses with this sub wavelength focussing property are referred to as 'perfect lenses'.

Remarkable as this lens is it suffers from some limitations: its properties depend on the condition $n = -1$ being met exactly and this can happen only at a single frequency because negative values of $\varepsilon, \mu$ imply dispersion. Furthermore the image is exactly the same size as the object: no magnification takes place. This latter limitation was partially lifted in [3, 4] where it was shown how to make a two dimensional version of the lens in the form of a cylinder with restricted magnifying powers. Here we show how to lift this restriction and design a spherical lens.

First we show how the concept of complementary media can be generalised from a simple slab to more complex situations. We ask the question: are there other instances of 'complementary media' one cancelling the optical effects of the other? This proves to be possible and is an essential first step in designing the perfect spherical lens.

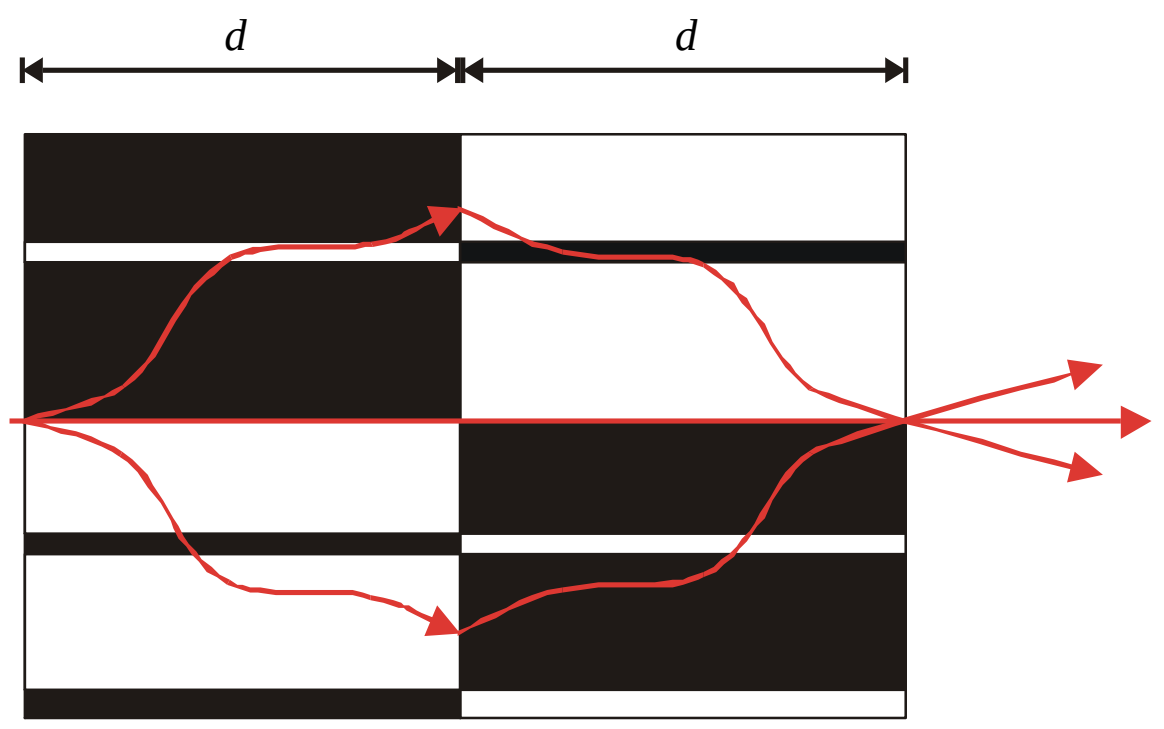

**Figure 2.** An alternative pair of complementary media, each cancelling the effect of the other. The light does not necessarily follow a straight line path in each medium. but the overall effect is as if a section of space thickness $2d$ were removed from the experiment.

Next we show that a square corner of $n=-1$ material has some characteristics in common with a corner reflector: it reverses the direction of light incident upon it. However it does more than a mirror can. It also reverses the direction of decay of the evanescent components of the light and is therefore 'perfect' in the same sense that the planar slab is a perfect lens. Something that is not possible with a corner mirror, is to combine to corners of negative material so that they share the same corner. Notomi has discussed this configuration and shown that classical rays radiating from an object in one of the quadrants are returned to the same spot. Here we show an even more powerful result: the returning image is again perfect. in that all components return, both radiative and evanescent.

Finally we remark that the technique we use in proving these results, the equivalence of coordinate transformations to a change in $\varepsilon, \mu$, is a powerful and widely applicable tool which we expect to show its value in generating even more results in the future.

## 2. A Generalised Lens Theorem

The original 'perfect lens' envisaged a slab of material with $\varepsilon=-1, \quad \mu=-1$. However focussing can be shown to occur under much more general conditions. We shall show that a system for which,

$$\begin{aligned} \varepsilon_1 &= +\varepsilon(x,y), \quad \mu_1 = +\mu(x,y), \quad -d<z<0 \\ \varepsilon_2 &= -\varepsilon(x,y), \quad \mu_2 = -\mu(x,y), \quad 0<z<d \end{aligned} \tag{2}$$

will show identical focussing. We need to show that the fields at $z=-d$ are identical to the fields at $z=+d$ under all circumstances. We shall first do this for a single Bloch wave, and then generalise by recognising that any field can be written as a sum over Bloch waves. Note that we have not defined the medium outside the confines of $-d<z<+d$ as there is no need. Focussing takes place irrespective of the medium in which the lens is embedded. The theorem shows that optically speaking the medium $-d<z<+d$ behaves as though it had zero thickness, as if it were removed and the two remaining halves moved together to close the gap.

We add a caveat about negative materials. A negative value of $\varepsilon$ necessarily implies dispersion as Smith and Kroll [5] have shown,

$$\frac{\partial(\omega\varepsilon)}{\partial\omega} > 0 \tag{3}$$

which in turn implies that we must be careful about causality which requires that calculations are always make including a small imaginary part to $\omega$, $i\delta$, and then the limit $\delta \to 0$ taken. Since $\varepsilon$ is dispersive, this implies that calculations must always be made with a small absorptive imaginary part to $\varepsilon$. In particular the case of $\varepsilon = \mu = -1$ is highly singular and erroneous results may follow if this limiting process is not employed. In all that follows the statement $\varepsilon = \mu = -1$ must always be understood as,

$$\lim_{\delta\to 0} \varepsilon(\omega + i\delta) = -1, \quad \lim_{\delta\to 0} \mu(\omega + i\delta) = -1 \tag{4}$$

We start from Maxwell's equations,

$$\nabla \times \mathbf{E} = +i\omega\mu(x,y)\mu_0\mathbf{H}, \quad \nabla \times \mathbf{H} = -i\omega\varepsilon(x,y)\varepsilon_0\mathbf{E} \tag{5}$$

and decompose the fields into Fourier components,

$$\begin{aligned}
\mathbf{E}_1(x,y,z) &= \exp(+ik_{1z}z) \sum_{k_x,k_y} \mathbf{E}_1(k_x,k_y)\exp(ik_x x + ik_y y), \quad -d < z < 0 \\
\mathbf{E}_2(x,y,z) &= \exp(+ik_{2z}z) \sum_{k_x,k_y} \mathbf{E}_2(k_x,k_y)\exp(ik_x x + ik_y y), \quad 0 < z < d
\end{aligned} \tag{6}$$

where we have assumed the Bloch theorem in the region $-d < z < 0$, but need to prove that the proposed solution meets the boundary conditions at $z = 0$.

We know that this solution obeys Maxwell's equations ,

$$\begin{aligned}
& k_{1z}\hat{\mathbf{z}} \times \left[E_{1x}(k_x,k_y)\hat{\mathbf{x}} + E_{1y}(k_x,k_y)\hat{\mathbf{y}}\right] + (k_x\hat{\mathbf{x}} + k_y\hat{\mathbf{y}}) \times E_{1z}(k_x,k_y)\hat{\mathbf{z}} \\
& \quad = +\omega\mu_0 \sum_{k_x',k_y'} \mu_1(k_x,k_y;k_x',k_y')\left[H_{1x}(k_x',k_y')\hat{\mathbf{x}} + H_{1y}(k_x',k_y')\hat{\mathbf{y}}\right] \\
& (k_x\hat{\mathbf{x}} + k_y\hat{\mathbf{y}}) \times \left[E_{1x}(k_x,k_y)\hat{\mathbf{x}} + E_{1y}(k_x,k_y)\hat{\mathbf{y}}\right] \\
& \quad = +\omega\mu_0 \sum_{k_x',k_y'} \mu_1(k_x,k_y;k_x',k_y') H_{1z}(k_x',k_y')\hat{\mathbf{z}} \\
& k_{1z}\hat{\mathbf{z}} \times \left[H_{1x}(k_x,k_y)\hat{\mathbf{x}} + H_{1y}(k_x,k_y)\hat{\mathbf{y}}\right] + (k_x\hat{\mathbf{x}} + k_y\hat{\mathbf{y}}) \times H_{1z}(k_x,k_y)\hat{\mathbf{z}} \\
& \quad = -\omega\varepsilon_0 \sum_{k_x',k_y'} \varepsilon_1(k_x,k_y;k_x',k_y')\left[E_{1x}(k_x',k_y')\hat{\mathbf{x}} + E_{1y}(k_x',k_y')\hat{\mathbf{y}}\right] \\
& (k_x\hat{\mathbf{x}} + k_y\hat{\mathbf{y}}) \times \left[H_{1x}(k_x,k_y)\hat{\mathbf{x}} + H_{1y}(k_x,k_y)\hat{\mathbf{y}}\right] \\
& \quad = -\omega\varepsilon_0 \sum_{k_x',k_y'} \varepsilon_1(k_x,k_y;k_x',k_y') E_{1z}(k_x',k_y')\hat{\mathbf{z}}
\end{aligned} \tag{7}$$

Next substitute,

$$E_{2x}\left(k_x,k_y\right)=E_{1x}\left(k_x,k_y\right),\quad E_{2y}\left(k_x,k_y\right)=E_{1y}\left(k_x,k_y\right),\quad E_{2z}\left(k_x,k_y\right)=-E_{1z}\left(k_x,k_y\right),$$
$$H_{2x}\left(k_x,k_y\right)=H_{1x}\left(k_x,k_y\right),\quad H_{2y}\left(k_x,k_y\right)=H_{1y}\left(k_x,k_y\right),\quad H_{2z}\left(k_x,k_y\right)=-H_{1z}\left(k_x,k_y\right),$$
$$k_{2z}=-k_{1z},\quad \varepsilon_2\left(k_x,k_y;k_x',k_y'\right)=-\varepsilon_1\left(k_x,k_y;k_x',k_y'\right),\quad \mu_2\left(k_x,k_y;k_x',k_y'\right)=-\mu_1\left(k_x,k_y;k_x',k_y'\right)$$

and it follows that,

$$\begin{aligned}
&-k_{2z}\hat{\mathbf{z}}\times\left[E_{2x}\left(k_x,k_y\right)\hat{\mathbf{x}}+E_{2y}\left(k_x,k_y\right)\hat{\mathbf{y}}\right]-\left(k_x\hat{\mathbf{x}}+k_y\hat{\mathbf{y}}\right)\times E_{2z}\left(k_x,k_y\right)\hat{\mathbf{z}}\\
&\qquad=-\omega\mu_0\sum_{k_x',k_y'}\mu_2\left(k_x,k_y;k_x',k_y'\right)\left[H_{2x}\left(k_x',k_y'\right)\hat{\mathbf{x}}+H_{2y}\left(k_x',k_y'\right)\hat{\mathbf{y}}\right]\\
&\left(k_x\hat{\mathbf{x}}+k_y\hat{\mathbf{y}}\right)\times\left[E_{2x}\left(k_x,k_y\right)\hat{\mathbf{x}}+E_{2y}\left(k_x,k_y\right)\hat{\mathbf{y}}\right]\\
&\qquad=+\omega\mu_0\sum_{k_x',k_y'}\mu_2\left(k_x,k_y;k_x',k_y'\right)H_{2z}\left(k_x',k_y'\right)\hat{\mathbf{z}}\\
&-k_{2z}\hat{\mathbf{z}}\times\left[H_{2x}\left(k_x,k_y\right)\hat{\mathbf{x}}+H_{2y}\left(k_x,k_y\right)\hat{\mathbf{y}}\right]-\left(k_x\hat{\mathbf{x}}+k_y\hat{\mathbf{y}}\right)\times H_{2z}\left(k_x,k_y\right)\hat{\mathbf{z}}\\
&\qquad=+\omega\varepsilon_0\sum_{k_x',k_y'}\varepsilon_2\left(k_x,k_y;k_x',k_y'\right)\left[E_{2x}\left(k_x',k_y'\right)\hat{\mathbf{x}}+E_{2y}\left(k_x',k_y'\right)\hat{\mathbf{y}}\right]\\
&\left(k_x\hat{\mathbf{x}}+k_y\hat{\mathbf{y}}\right)\times\left[H_{2x}\left(k_x,k_y\right)\hat{\mathbf{x}}+H_{2y}\left(k_x,k_y\right)\hat{\mathbf{y}}\right]\\
&\qquad=-\omega\varepsilon_0\sum_{k_x',k_y'}\varepsilon_2\left(k_x,k_y;k_x',k_y'\right)E_{2z}\left(k_x',k_y'\right)\hat{\mathbf{z}}
\end{aligned}\tag{9}$$

which rearranges to,

$$\begin{aligned}
&+k_{2z}\hat{\mathbf{z}}\times\left[E_{2x}\left(k_x,k_y\right)\hat{\mathbf{x}}+E_{2y}\left(k_x,k_y\right)\hat{\mathbf{y}}\right]+\left(k_x\hat{\mathbf{x}}+k_y\hat{\mathbf{y}}\right)\times E_{2z}\left(k_x,k_y\right)\hat{\mathbf{z}}\\
&\qquad=+\omega\mu_0\sum_{k_x',k_y'}\mu_2\left(k_x,k_y;k_x',k_y'\right)\left[H_{2x}\left(k_x',k_y'\right)\hat{\mathbf{x}}+H_{2y}\left(k_x',k_y'\right)\hat{\mathbf{y}}\right]\\
&\left(k_x\hat{\mathbf{x}}+k_y\hat{\mathbf{y}}\right)\times\left[E_{2x}\left(k_x,k_y\right)\hat{\mathbf{x}}+E_{2y}\left(k_x,k_y\right)\hat{\mathbf{y}}\right]\\
&\qquad=+\omega\mu_0\sum_{k_x',k_y'}\mu_2\left(k_x,k_y;k_x',k_y'\right)H_{2z}\left(k_x',k_y'\right)\hat{\mathbf{z}}\\
&+k_{2z}\hat{\mathbf{z}}\times\left[H_{2x}\left(k_x,k_y\right)\hat{\mathbf{x}}+H_{2y}\left(k_x,k_y\right)\hat{\mathbf{y}}\right]+\left(k_x\hat{\mathbf{x}}+k_y\hat{\mathbf{y}}\right)\times H_{2z}\left(k_x,k_y\right)\hat{\mathbf{z}}\\
&\qquad=-\omega\varepsilon_0\sum_{k_x',k_y'}\varepsilon_2\left(k_x,k_y;k_x',k_y'\right)\left[E_{2x}\left(k_x',k_y'\right)\hat{\mathbf{x}}+E_{2y}\left(k_x',k_y'\right)\hat{\mathbf{y}}\right]\\
&\left(k_x\hat{\mathbf{x}}+k_y\hat{\mathbf{y}}\right)\times\left[H_{2x}\left(k_x,k_y\right)\hat{\mathbf{x}}+H_{2y}\left(k_x,k_y\right)\hat{\mathbf{y}}\right]\\
&\qquad=-\omega\varepsilon_0\sum_{k_x',k_y'}\varepsilon_2\left(k_x,k_y;k_x',k_y'\right)E_{2z}\left(k_x',k_y'\right)\hat{\mathbf{z}}
\end{aligned}\tag{10}$$

i.e. the new fields solve Maxwell's equations for the new $\varepsilon_2, \mu_2$. The fields in the two regions also match across the boundary since the parallel components of **E** and **H** are defined to be continuous, and the *z* components equal and opposite. Therefore we have,

$$\mathbf{E}(x, y, z=+a) = \mathbf{E}(x, y)\exp(-ik_z a) = \mathbf{E}(x, y, z=-a), \qquad a>0 \tag{11}$$

and therefore the fields to the left and right are repeated in inverse order exactly as was the case with our original uniform slab.

In general the fields will be sums over $k_z$, but the result holds for each $k_z$ separately and therefore for any summation.

The result can be generalised to the case where $\varepsilon, \mu$ are tensors,

$$\varepsilon = \begin{bmatrix} \varepsilon_{xx} & \varepsilon_{xy} & \varepsilon_{xz} \\ \varepsilon_{yx} & \varepsilon_{yy} & \varepsilon_{yz} \\ \varepsilon_{zx} & \varepsilon_{zy} & \varepsilon_{zz} \end{bmatrix}, \quad \mu = \begin{bmatrix} \mu_{xx} & \mu_{xy} & \mu_{xz} \\ \mu_{yx} & \mu_{yy} & \mu_{yz} \\ \mu_{zx} & \mu_{zy} & \mu_{zz} \end{bmatrix} \tag{12}$$

except that now the complementary media have the form,

$$\begin{gathered}
\varepsilon_1 = \begin{bmatrix} \varepsilon_{xx} & \varepsilon_{xy} & \varepsilon_{xz} \\ \varepsilon_{yx} & \varepsilon_{yy} & \varepsilon_{yz} \\ \varepsilon_{zx} & \varepsilon_{zy} & \varepsilon_{zz} \end{bmatrix}, \quad \mu_1 = \begin{bmatrix} \mu_{xx} & \mu_{xy} & \mu_{xz} \\ \mu_{yx} & \mu_{yy} & \mu_{yz} \\ \mu_{zx} & \mu_{zy} & \mu_{zz} \end{bmatrix}, \qquad -d<z<0 \\
\varepsilon_2 = \begin{bmatrix} -\varepsilon_{xx} & -\varepsilon_{xy} & +\varepsilon_{xz} \\ -\varepsilon_{yx} & -\varepsilon_{yy} & +\varepsilon_{yz} \\ +\varepsilon_{zx} & +\varepsilon_{zy} & -\varepsilon_{zz} \end{bmatrix}, \quad \mu_2 = \begin{bmatrix} -\mu_{xx} & -\mu_{xy} & +\mu_{xz} \\ -\mu_{yx} & -\mu_{yy} & +\mu_{yz} \\ +\mu_{zx} & +\mu_{zy} & -\mu_{zz} \end{bmatrix}, \quad 0<z<+d
\end{gathered} \tag{13}$$

We can express our theorem in a graphical fashion: two complementary media have an optical sum of zero. We calculate the optical response of the rest of the system outside the lens region by cutting out the complementary media which comprise the lens and closing the gap between the two remaining halves of the systems.

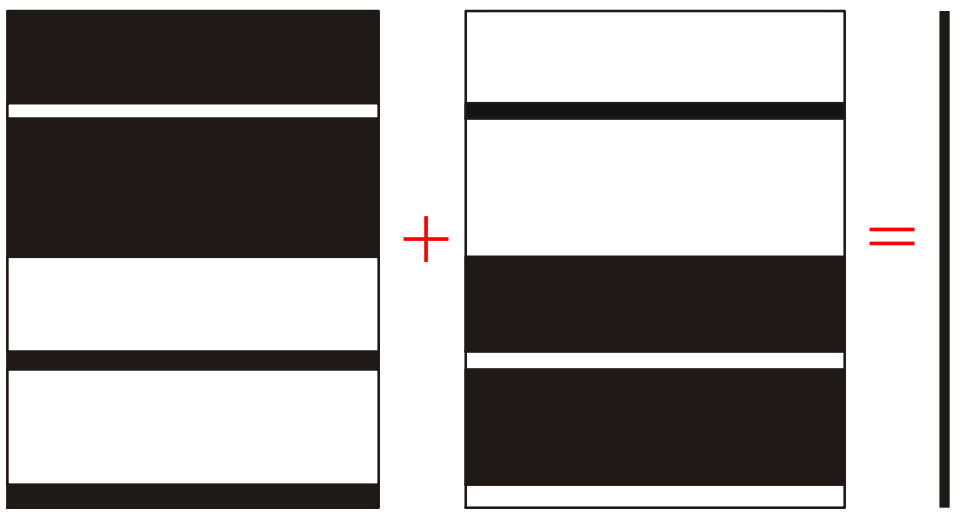

**Figure 3.** A graphical expression of our new theorem: complementary halves sum to zero. The optical properties of the rest of the system can be calculated by cutting out the media and closing the gap.

This enables us to go one step further with the proof. Consider a new system in which,

$$\begin{aligned}
\varepsilon_1 &= +\varepsilon(x,y), & \mu_1 &= +\mu(x,y), & -f<z<-d \\
\varepsilon_2 &= +\varepsilon'(x,y), & \mu_2 &= +\mu'(x,y), & -d<z<0 \\
\varepsilon_3 &= -\varepsilon'(x,y), & \mu_3 &= -\mu'(x,y), & 0<z<+d \\
\varepsilon_4 &= -\varepsilon(x,y), & \mu_4 &= -\mu(x,y), & +d<z<+f
\end{aligned} \tag{14}$$

i.e. the media are piecewise complementary. We apply the lens theorem to the inner complementary pair and reduce to a new system,

$$\begin{aligned} \varepsilon_1 &= +\varepsilon(x,y), & \mu_1 &= +\mu(x,y), & -(f-d) &< z < 0 \\ \varepsilon_4 &= -\varepsilon(x,y), & \mu_4 &= -\mu(x,y), & 0 &< z < +(f-d) \end{aligned} \tag{15}$$

and as a final step, eliminate the outer pair of complementary media. The process is demonstrated graphically in figure 4.

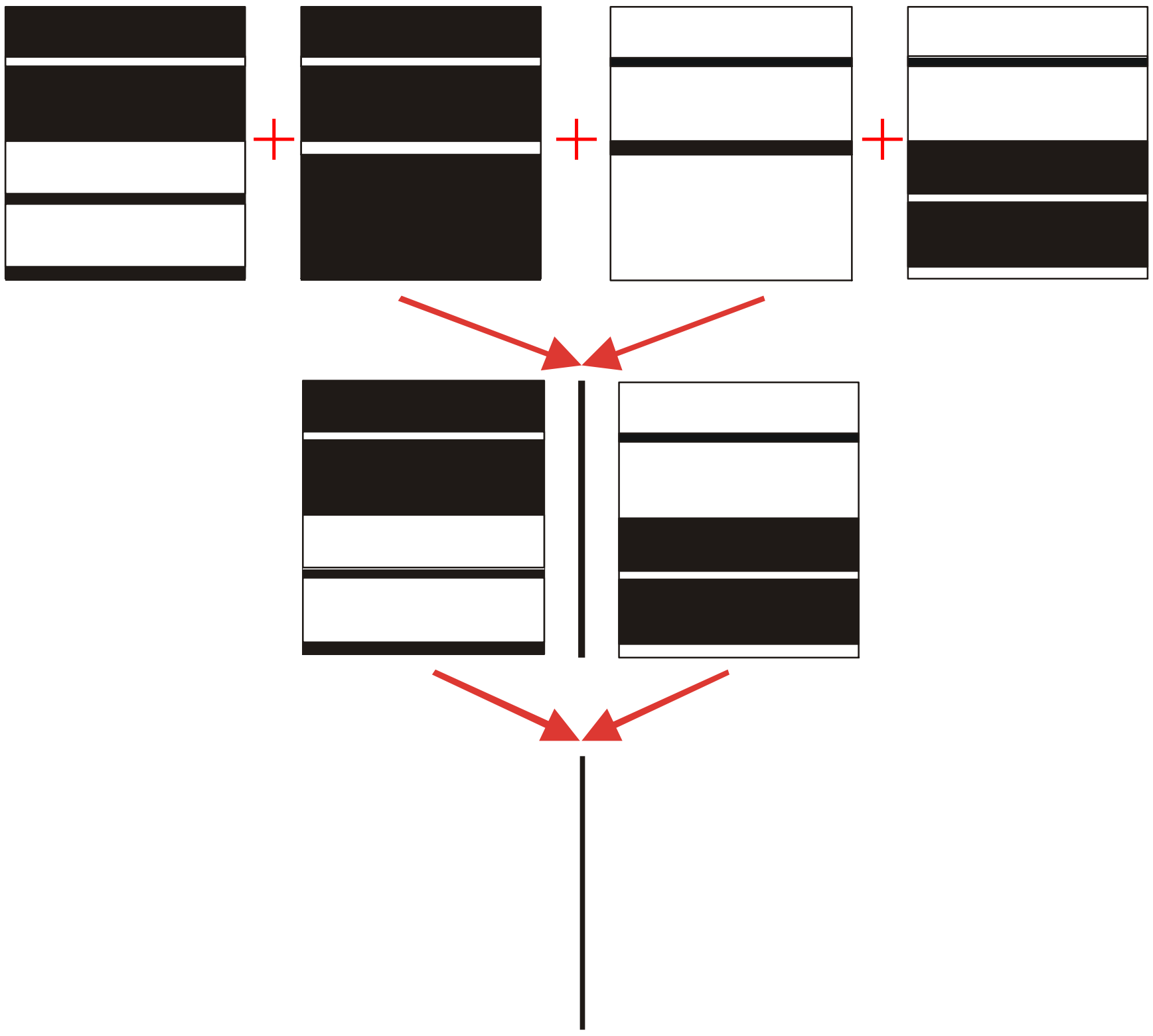

**Figure 4.** Two sets of optically complementary media can be eliminated in a pair wise fashion simply by applying twice the process shown in figure 3. Obviously this process can be generalised to any number of complementary pairs.

As a result we may generalise the lens theorem even further to any region with the property,

$$\begin{aligned} \varepsilon_1 &= +\varepsilon(x,y,z), & \mu_1 &= +\mu(x,y,z), & -d &< z < 0 \\ \varepsilon_2 &= -\varepsilon(x,y,-z), & \mu_2 &= -\mu(x,y,-z), & 0 &< z < d \end{aligned} \tag{16}$$

In other words to any region that is mirror antisymmetric about a plane. Such regions can be optically eliminated from the system.

Other even more complex systems can be envisaged for which a proof of vanishing can be concocted, but we leave these as an exercise for the reader.

## 3. Some Examples of Complementary Media

To verify the generalised perfect lens theorem above, we carried out numerical simulations of media where $\varepsilon, \mu$ vary in the $xy$ plane taken to be is transverse to the lens axis. In the first set of calculations each section of the medium is piecewise translationally invariant along the axis of the lens, the $z$ direction. In the second set this restriction is relaxed and we treat the more general case of reflections antisymmetry.

The first set of calculations assume $\varepsilon, \mu$ periodic on the *xy* plane defined as follows,

$$\begin{aligned}
\varepsilon &= -\varepsilon_1 - \varepsilon_2 \sin^2(2\pi x/\lambda_\varepsilon), \quad && |z| < d \\
\varepsilon &= +\varepsilon_1 + \varepsilon_2 \sin^2(2\pi x/\lambda_\varepsilon), \quad && d < |z| < 2d \\
\mu &= -\mu_1 - \mu_2 \sin^2(2\pi x/\lambda_\mu), \quad && |z| < d \\
\mu &= +\mu_1 + \mu_2 \sin^2(2\pi x/\lambda_\mu), \quad && d < |z| < 2d
\end{aligned} \tag{17}$$

For simplicity we choose the media to be periodic along the $x$ coordinate and invariant along the $y$ coordinate. Figure 5 illustrates the situation.

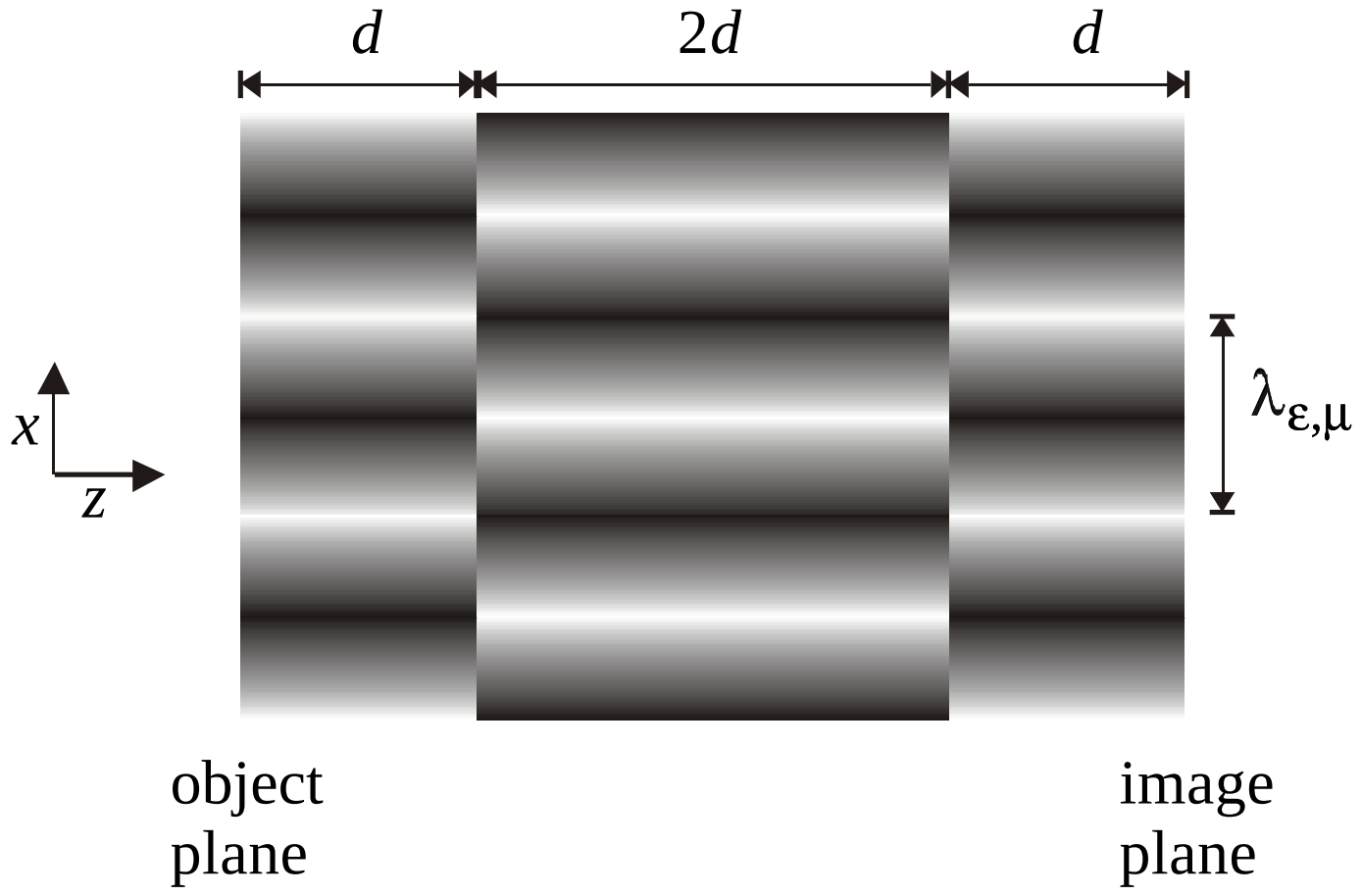

**Figure 5.** The geometry assumed for the numerical simulations on complementary media. The object and image planes are at $z = -2d$ and $z = +2d$ respectively in media that are periodic along the $x-$ axis. $\lambda_\varepsilon$ and $\lambda_\mu$ are the spatial periodicities for $\varepsilon, \mu$ respectively.

We calculate the transmission and reflection coefficients for plane waves incident on the system using the PHOTON codes [6] based on the transfer matrix method [7,8] developed earlier in our group. The transmittance, $|T|^2$, the reflectance, $|R|^2$, and the phase, $\arg T$, of the transmitted wave are calculated for both the S- and P-polarized waves as a function of the transverse wave-vector $k_x$. In Figure 6 we show the results for,

$$\varepsilon_1 = 0.5, \ \mu_1 = 0.5, \ \varepsilon_2 = 1.0, \ \mu_2 = 1.0, \ 2d = 0.45\mu\text{m} \tag{18}$$

where $\lambda$ is the wavelength of radiation in free space and $2d$ is the thickness of the slab. In Figure 6a and 6b, we show the results for $\lambda_\varepsilon = \lambda_\mu = 0.9\mu\text{m}, \ \lambda = 4.54\mu\text{m}$, and in Figures 6c and 6d, $\lambda_\varepsilon = 1.8\mu\text{m}, \ \lambda_\mu = 0.9\mu\text{m}, \ \lambda = 4.54\mu\text{m}$.

Our theory predicts that the complementary media should have unit transmission and zero reflection coefficients. Simulations will not reproduce these results exactly but only to within the accuracy of the simulations, as determined by the finite level of discretisation. Figure 6 shows that this is indeed the case for all wave vectors, both for propagating wave vectors, $k_x/k_0 < 1$, and evanescent waves, $k_x/k_0 > 1$. There are minor deviations of our calculations from the unit transmission coefficient for the perfect lens, particularly for wave vectors $k_x/k_0 \approx 1$.

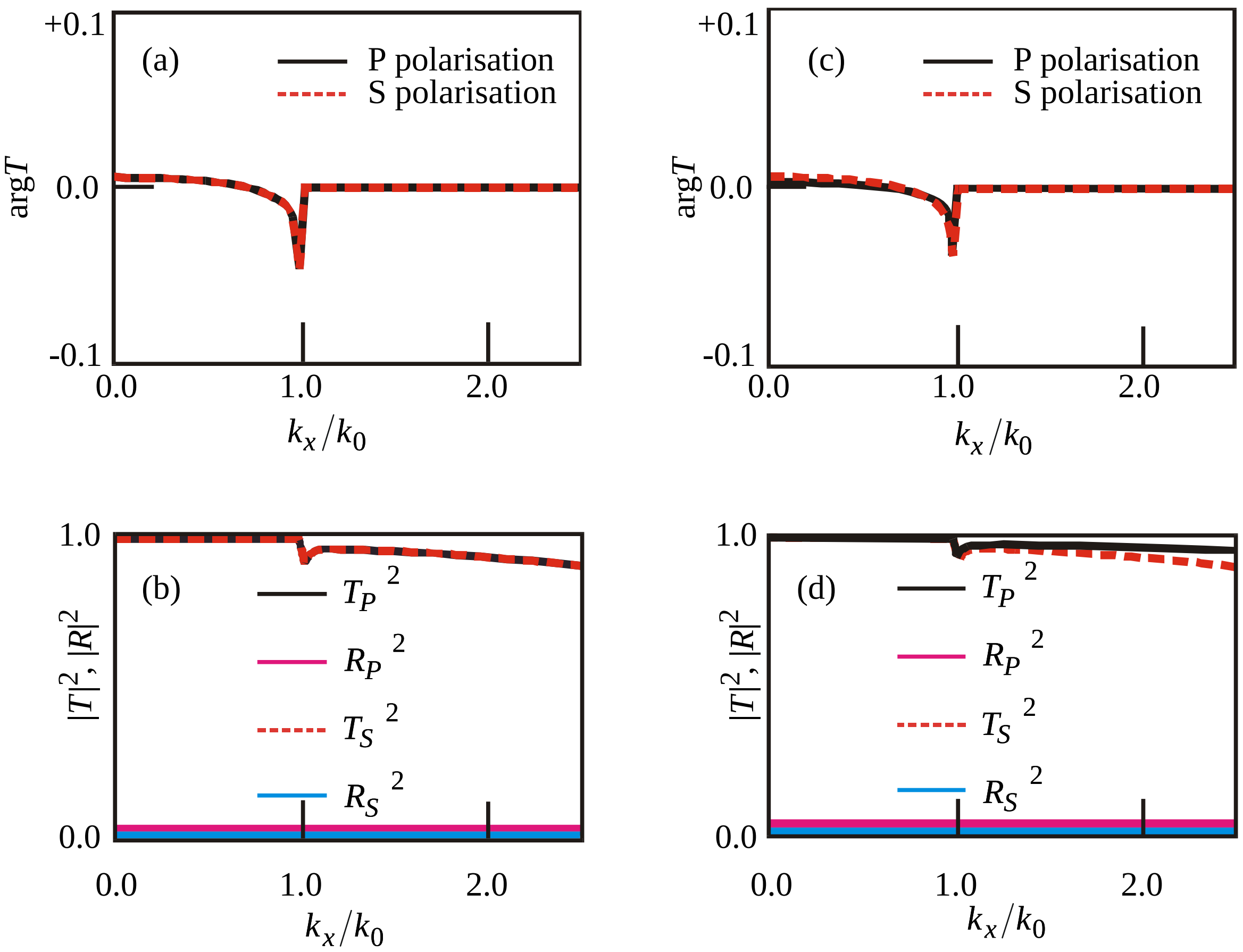

**Figure 6.** The phase of the transmitted wave $\arg T$, the transmittance $|T|^2$ and reflectance $|R|^2$ for the P- and S-polarised waves for the layered system of Figure 5, see equation (17). The figures on the left, (a) and (b) are for $\lambda_\varepsilon = \lambda_\mu = 0.9\mu\text{m}, \quad \lambda = 4.54\mu\text{m}$, (c) and (d) are for $\lambda_\varepsilon = 1.8\mu\text{m}, \ \lambda_\mu = 0.9\mu\text{m}, \quad \lambda = 4.54\mu\text{m}$.

In the next set of calculations we take a more complicated form of lens where the components are not translationally invariant in *z*, but instead obey the less restrictive condition of symmetry about $z = 0$. The test parameters are defined as follows:

$$
\begin{aligned}
\varepsilon &= -1.0 \qquad -\frac{|z|}{d}\sin^2\left(2\pi x/\lambda_\varepsilon\right), \qquad |z| < d \\
\varepsilon &= +1.0 + \frac{2d - |z|}{d}\sin^2\left(2\pi x/\lambda_\varepsilon\right), \quad d < |z| < 2d \\
\mu &= -1.0 \qquad -\frac{|z|}{d}\sin^2\left(2\pi x/\lambda_\mu\right), \qquad |z| < d \\
\mu &= +1.0 + \frac{2d - |z|}{d}\sin^2\left(2\pi x/\lambda_\mu\right), \quad d < |z| < 2d
\end{aligned}
\tag{19}
$$

Below in figure 7 we show the results. Deviations of the transmittance from unity, reflectance from zero and the phase of the transmitted wave at the image plane from zero are small: in this example discontinuities in the parameters are minimised in order to make the simulations more accurate.

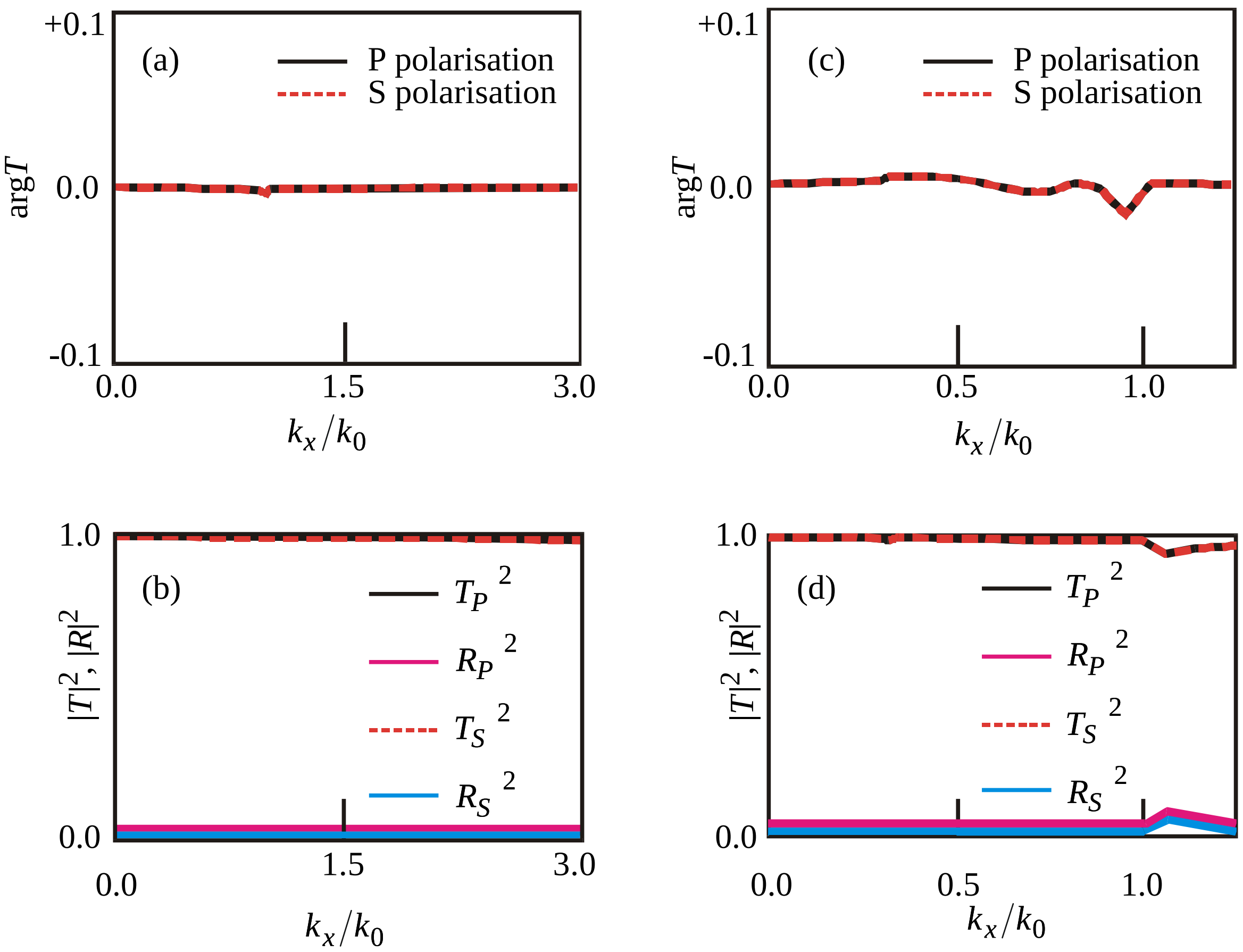

**Figure 7.** We show the phase of the transmitted wave, the transmittance $|T|^2$ and reflectance $|R|^2$ for the P- and S-polarised waves for the case defined by equation (19). (a) and (b) are for $\lambda_\varepsilon = \lambda_\mu = 0.9\mu\text{m}$, $\lambda = 4.54\mu m$ (low energy). (c) and (d) are for $\lambda_\varepsilon = \lambda_\mu = 1.8\mu\text{m}$ and $\lambda = 0.617\mu\text{m}$.

To summarise these calculations: to within the accuracy of our computations, determined by the level of discretisation, we offer clear numerical proof for the generalisation of the perfect lens theorem to media with transverse spatial variation. We have shown it to hold for periodic media with various arbitrarily chosen periodicities. It follows that it holds in the general case since any general transverse spatial variation can be Fourier decomposed into periodic components.

## 4. A Perfect Spherical Lens by Coordinate Transformation

In the previous sections we have shown how two complementary regions of space optically cancel: each region is the inverted mirror image of the other. Since the geometry is planar images are necessarily the same size as the object. To magnify we must work with curved surfaces. In the following sections we show how to make a further generalisation of the cancellation principle by using a coordinate transformation to distort the space. We give examples below of two spherical annuli which cancel to produced magnified images.

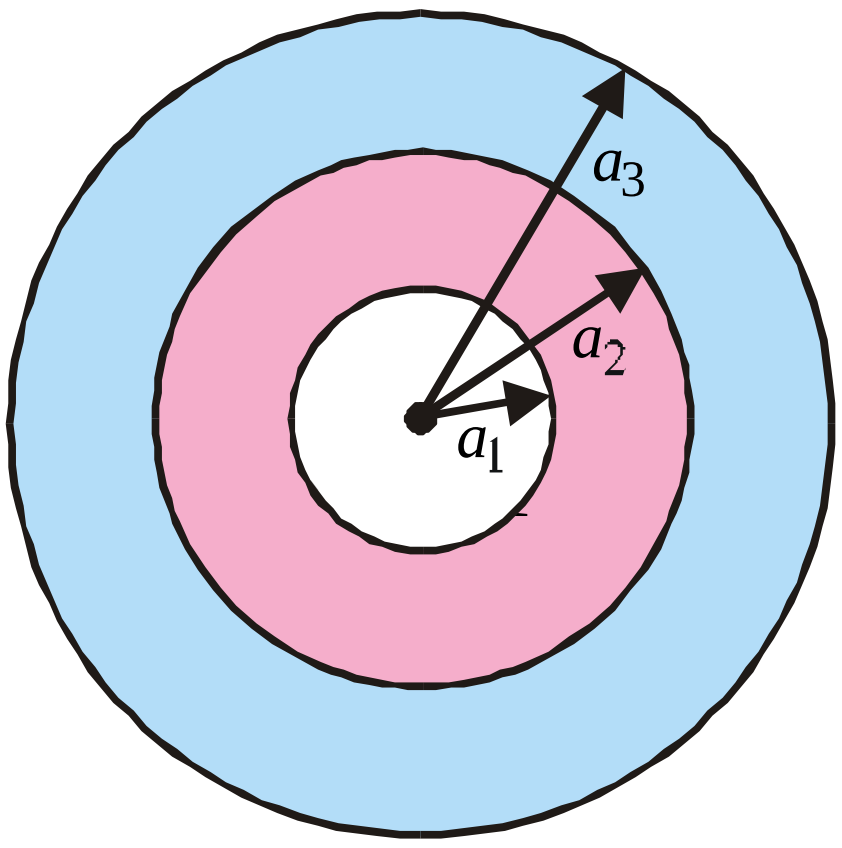

**Figure 8.** A design for a spherical lens. Objects on the surface $r = a_1$ are refocused and magnified on the outer surface $r = a_3$ provided that the space in between houses optically complementary media.

Consider a spherically symmetric system into which we wish to introduce a lens. The lens will comprise two parts, the material in $a_1 < r < a_2$ being complementary to the material $a_2 < r < a_3$ so that objects on the surface $r = a_3$ are focussed onto the surface $r = a_1$, but reduced in size by a factor $a_1/a_3$ . We shall now show how to specify these materials

We apply a coordinate transformation,

$$q_1(x,y,z), \quad q_2(x,y,z), \quad q_3(x,y,z) \tag{20}$$

so that in the new frame of generalised axes $q_1, q_2, q_3$ assumed orthogonal,

$$\tilde{\varepsilon}_i = \varepsilon_i \frac{Q_1 Q_2 Q_3}{Q_i^2}, \quad \tilde{\mu}_i = \mu_i \frac{Q_1 Q_2 Q_3}{Q_i^2} \tag{21}$$

$$\tilde{E}_i = Q_i E_i, \quad \tilde{H}_i = Q_i H_i \tag{22}$$

where,

$$Q_i^2 = \left(\frac{\partial x}{\partial q_i}\right)^2 + \left(\frac{\partial y}{\partial q_i}\right)^2 + \left(\frac{\partial z}{\partial q_i}\right)^2 \tag{23}$$

Let us define a set of spherical coordinates as follows,

$$x = r_0 e^{\ell/\ell_0} \sin\theta\cos\phi, \quad y = r_0 e^{\ell/\ell_0} \sin\theta\sin\phi, \quad z = r_0 e^{\ell/\ell_0} \cos\theta \tag{24}$$

so that,

$$\begin{aligned}
Q_\ell &= r_0/\ell_0\, e^{\ell/\ell_0} \sqrt{\sin^2\theta\cos^2\phi + \sin^2\theta\sin^2\phi + \cos^2\theta} = r_0/\ell_0\, e^{\ell/\ell_0} \\
Q_\theta &= r_0 e^{\ell/\ell_0} \sqrt{\cos^2\theta\cos^2\phi + \cos^2\theta\sin^2\phi + \sin^2\theta} = r_0 e^{\ell/\ell_0} \\
Q_\phi &= r_0 e^{\ell/\ell_0} \sqrt{\sin^2\theta\sin^2\phi + \sin^2\theta\cos^2\phi} = r_0 e^{\ell/\ell_0} \sin\theta
\end{aligned} \tag{25}$$

$$Q_\ell Q_\theta Q_\phi = r_0^3/\ell_0\, e^{3\ell/\ell_0} \sin\theta$$

and from (21), (25)

$$\begin{aligned}
\tilde{\varepsilon}_\ell &= r_0 \ell_0 e^{\ell/\ell_0} \sin\theta\, \varepsilon_\ell, \quad \tilde{\varepsilon}_\theta = \frac{r_0 e^{\ell/\ell_0}}{\ell_0} \sin\theta\ \varepsilon_\theta, \quad \tilde{\varepsilon}_\phi = \frac{r_0 e^{\ell/\ell_0}}{\ell_0 \sin\theta} \varepsilon_\phi \\
\tilde{\mu}_\ell &= r_0 \ell_0 e^{\ell/\ell_0} \sin\theta\, \mu_\ell, \quad \tilde{\mu}_\theta = \frac{r_0 e^{\ell/\ell_0}}{\ell_0} \sin\theta\ \mu_\theta, \quad \tilde{\mu}_\phi = \frac{r_0 e^{\ell/\ell_0}}{\ell_0 \sin\theta} \varepsilon_\phi
\end{aligned} \tag{26}$$

where the $\ell$ coordinate is oriented along the radial direction. Now if we make the choice of $\ell_0 = 1$ and,

$$\begin{aligned}
\varepsilon_\ell &= \varepsilon_\theta = \varepsilon_\phi = \varepsilon e^{-\ell/\ell_0} = \varepsilon\, r_0/r \\
\mu_\ell &= \mu_\theta = \mu_\phi = \mu e^{-\ell/\ell_0} = \mu\, r_0/r
\end{aligned} \tag{27}$$

then the radial dependence goes away to give,

$$\begin{aligned}
\tilde{\varepsilon}_\ell &= \varepsilon r_0 \sin\theta, \quad \tilde{\varepsilon}_\theta = \varepsilon r_0 \sin\theta, \quad \tilde{\varepsilon}_\phi = \varepsilon \frac{r_0}{\sin\theta} \\
\tilde{\mu}_\ell &= \mu r_0 \sin\theta, \quad \tilde{\mu}_\theta = \mu r_0 \sin\theta, \quad \tilde{\mu}_\phi = \mu \frac{r_0}{\sin\theta}
\end{aligned} \tag{28}$$

In other words we have a system which is translationally invariant in one of the coordinates, $\ell$. If we insert two slabs of material complementary in the sense of equations (2), the first between,

$$\ln(a_2/r_0) = \ell_2 < \ell < \ell_3 = \ln(a_3/r_0) \tag{29}$$

the second between,

$$\ln(a_1/r_0) = \ell_1 < \ell < \ell_2 = \ln(a_2/r_0) \tag{30}$$

with the requirement that,

$$\ell_3 - \ell_2 = \ell_2 - \ell_1 \tag{31}$$

i.e.,

$$a_3 = a_2^2 / a_1 \tag{32}$$

Or specifically,

$$\begin{aligned}
\varepsilon_1 &= +\varepsilon(\theta,\phi)\frac{r_0}{r}, \quad \mu_1 = +\mu(\theta,\phi)\frac{r_0}{r}, \quad a_1 < r < a_2 \\
\varepsilon_2 &= -\varepsilon(\theta,\phi)\frac{r_0}{r}, \quad \mu_2 = -\mu(\theta,\phi)\frac{r_0}{r}, \quad a_2 < r < a_3 = a_2^2/a_1
\end{aligned} \tag{33}$$

Then an object located at $\ell_1$ will form an image at $\ell_3$, or viewed in the original coordinate frame, objects on the surface,

$$r = a_1 \tag{34}$$

will be reproduced as a perfect image on the surface,

$$r = a_3 = a_2^2 / a_1 \tag{35}$$

but magnified by a factor $a_2^2 / a_1^2$ .

We have described the region $a_1 < r < a_3$ but what of the other regions? In the $\ell\theta\phi$ frame it is clear what is meant when we say that the region $\ell_1 < \ell < \ell_3$ 'vanishes' from an optical standpoint: the optical properties of the system viewed from a point external to $\ell_1 < \ell < \ell_3$ can be calculated by cutting out the vanishing region and closing up the gap. See figure 2. How is this viewed in the old coordinate frame where the region of space to be eliminated lies between two spherical shells?

Let is start with the appearance of the system from outside, $r > a_3$. In $\ell\theta\phi$ we eliminate the lens region to give an equivalent optical system and we do this by moving all material for $\ell < \ell_1$ a distance $\ell_3 - \ell_1$ along the $\ell$ axis. On transforming back to the $xyz$ frame the values of $\varepsilon, \mu$ also change. For the outer region the equivalent $\varepsilon, \mu$ are given by,

$$\begin{aligned} \varepsilon_{eq}^{+}(r) &= \varepsilon(r), \quad r > a_3, \\ \mu_{eq}^{+}(r) &= \mu(r), \quad r > a_3 \end{aligned} \tag{36}$$

i.e. we have simply undone the original transformation. However the shifted material does not go back into the same place it did before and therefore the transformation which depends on the radius, is slightly different,

$$\begin{aligned} \varepsilon_{eq}^{+}(r) &= \frac{a_1}{a_3}\varepsilon\left(\frac{a_1}{a_3}r\right) = \frac{a_1^2}{a_2^2}\varepsilon\left(\frac{a_1^2}{a_2^2}r\right), \quad r < a_3, \\ \mu_{eq}^{+}(r) &= \frac{a_1}{a_3}\mu\left(\frac{a_1}{a_3}r\right) = \frac{a_1^2}{a_2^2}\mu\left(\frac{a_1^2}{a_2^2}r\right), \quad r < a_3 \end{aligned} \tag{37}$$

i.e. as well as shifting the material, we also have to renormalise the magnitude of the $\varepsilon, \mu$. Conversely if we make an optical experiment inside the lensing region the optically equivalent system is given by,

$$\begin{aligned} \varepsilon_{eq}^{-}(r) &= \varepsilon(r), \quad r < a_1, \\ \mu_{eq}^{-}(r) &= \mu(r), \quad r < a_1 \end{aligned} \tag{38}$$

and,

$$\begin{aligned} \varepsilon_{eq}^{-}(r) &= \frac{a_3}{a_1}\varepsilon\left(\frac{a_3}{a_1}r\right) = \frac{a_2^2}{a_1^2}\varepsilon\left(\frac{a_2^2}{a_1^2}r\right), \quad r > a_1, \\ \mu_{eq}^{-}(r) &= \frac{a_3}{a_1}\mu\left(\frac{a_3}{a_1}r\right) = \frac{a_2^2}{a_1^2}\mu\left(\frac{a_2^2}{a_1^2}r\right), \quad r > a_1, \end{aligned} \tag{39}$$

These transformations are required thermodynamically because we cannot make a perfect magnified image in the same material as the source, as that would be equivalent to altering the temperature through an optical trick. Even negative materials cannot violate thermodynamics.

## 5. A Conventional Proof of the Spherical Lens

We give here a more conventional proof of the spherical lens. For simplicity we give the proof under more restricted conditions that those above. Let us assume that the system is spherically symmetric and that $\varepsilon(r) = \mu(r)$. We shall solve for the radial components of the electric and magnetic fields as follows.

$$\nabla\times\mathbf{E} = +i\omega\mu_0\mu(r)\mathbf{H}, \quad \nabla\times\mathbf{H} = -i\omega\varepsilon_0\varepsilon(r)\mathbf{E} \tag{40}$$

and hence from the Appendix equation (74),

$$\begin{aligned}(\nabla\times\nabla\times\mathbf{E})_r &= -\frac{1}{r^3}\frac{\partial}{\partial r}\left(r^2\frac{\partial(rE_r)}{\partial r}\right) - \frac{2}{r}\frac{\varepsilon'(r)}{\varepsilon(r)}E_r - \frac{\partial}{\partial r}\left[\frac{\varepsilon'(r)}{\varepsilon(r)}E_r\right] - \nabla^2_{\theta\phi}E_r \\ &= +\omega^2 c_0^{-2}\varepsilon(r)\mu(r)E_r\end{aligned} \tag{74}$$

which is an equation involving only the $rth$ component of $\mathbf{E}$ and describes a transverse magnetic (TM) field An analogous TE equation controls the behaviour of $H_r$ so that the arguments below can be repeated for the magnetic field component.

Substituting the following trial solution,

$$E_r = nr^{n-1}Y_{\ell m}(\theta\phi), \quad \varepsilon(r) = \mu(r) = \alpha r^p \tag{41}$$

into equation (74) gives,

$$\omega^2 c_0^{-2}\alpha^2 r^{2p+2} = -n(n+1+p) + \ell(\ell+1) \tag{42}$$

From the left hand side we require that

$$p = -1 \tag{43}$$

and from the right hand side,

$$n = \pm\sqrt{\ell(\ell+1) - \omega^2 c_0^{-2}\alpha^2} \tag{44}$$

Therefore let us define a system in which,

$$\begin{aligned}
\varepsilon_1 &= +\varepsilon\frac{r_0}{r}, \quad \mu_1 = +\mu\frac{r_0}{r}, \quad 0 < r < a_1 \\
\varepsilon_2 &= +\varepsilon\frac{r_0}{r}, \quad \mu_1 = +\mu\frac{r_0}{r}, \quad a_1 < r < a_2 \\
\varepsilon_3 &= -\varepsilon\frac{r_0}{r}, \quad \mu_2 = -\mu\frac{r_0}{r}, \quad a_2 < r < a_3 \\
\varepsilon_4 &= +\varepsilon\frac{r_0}{r}, \quad \mu_1 = +\mu\frac{r_0}{r}, \quad a_3 < r < \infty
\end{aligned} \tag{45}$$

which is a simplified version of the lens prescribed in section 4. The solutions for the radial fields are given by,

$$\begin{aligned}
E_r &= +[r]^{+\sqrt{\ell(\ell+1)-\omega^2 c_0^{-2}\alpha^2}-1}Y_{\ell m}(\theta\phi), \quad 0 < r < a_2 \\
E_r &= -\left[a_2^2\right]^{+\sqrt{\ell(\ell+1)-\omega^2 c_0^{-2}\alpha^2}}[r]^{-\sqrt{\ell(\ell+1)-\omega^2 c_0^{-2}\alpha^2}-1}Y_{\ell m}(\theta\phi), \quad a_2 < r < a_3 \\
E_r &= +\left[\frac{a_2^2}{a_3^2}\right]^{+\sqrt{\ell(\ell+1)-\omega^2 c_0^{-2}\alpha^2}}[r]^{+\sqrt{\ell(\ell+1)-\omega^2 c_0^{-2}\alpha^2}-1}Y_{\ell m}(\theta\phi), \quad a_3 < r < \infty
\end{aligned} \tag{46}$$

This solution matches the fields at the boundaries. It fulfils the function of a lens as we can see by rewriting,

$$E_r = +r^{-1}\left[\frac{a_2^2}{a_3^2} r\right]^{+\sqrt{\ell(\ell+1)-\omega^2 c_0^{-2}\alpha^2}} Y_{\ell m}(\theta\phi), \quad a_3 < r < \infty \tag{47}$$

so that, apart from a scaling factor of $r$, the fields on the surface of the sphere $r = a_3$ are identical with those on the sphere $r = a_1 = a_2^2 / a_3$.

Notice how much more powerful and elegant is the proof by coordinate transformation.

## 6. Negative Corners: a Perfect Corner Reflector by Coordinate Transformation

Material with refractive index $\lim n \to -1$ has some things in common with a mirror: the angle of refraction inside the medium is equal to the incident angle, just as a ray is reflected from a mirror with equal incident and reflected angles. One of the tricks to play with mirrors is the corner reflector which has the property that all incident rays emerge from the corner with the *xy* components of their wave vectors reversed, assuming the corner is bent in the *xy* plane. A ray construction, figure 9, shows that this is a

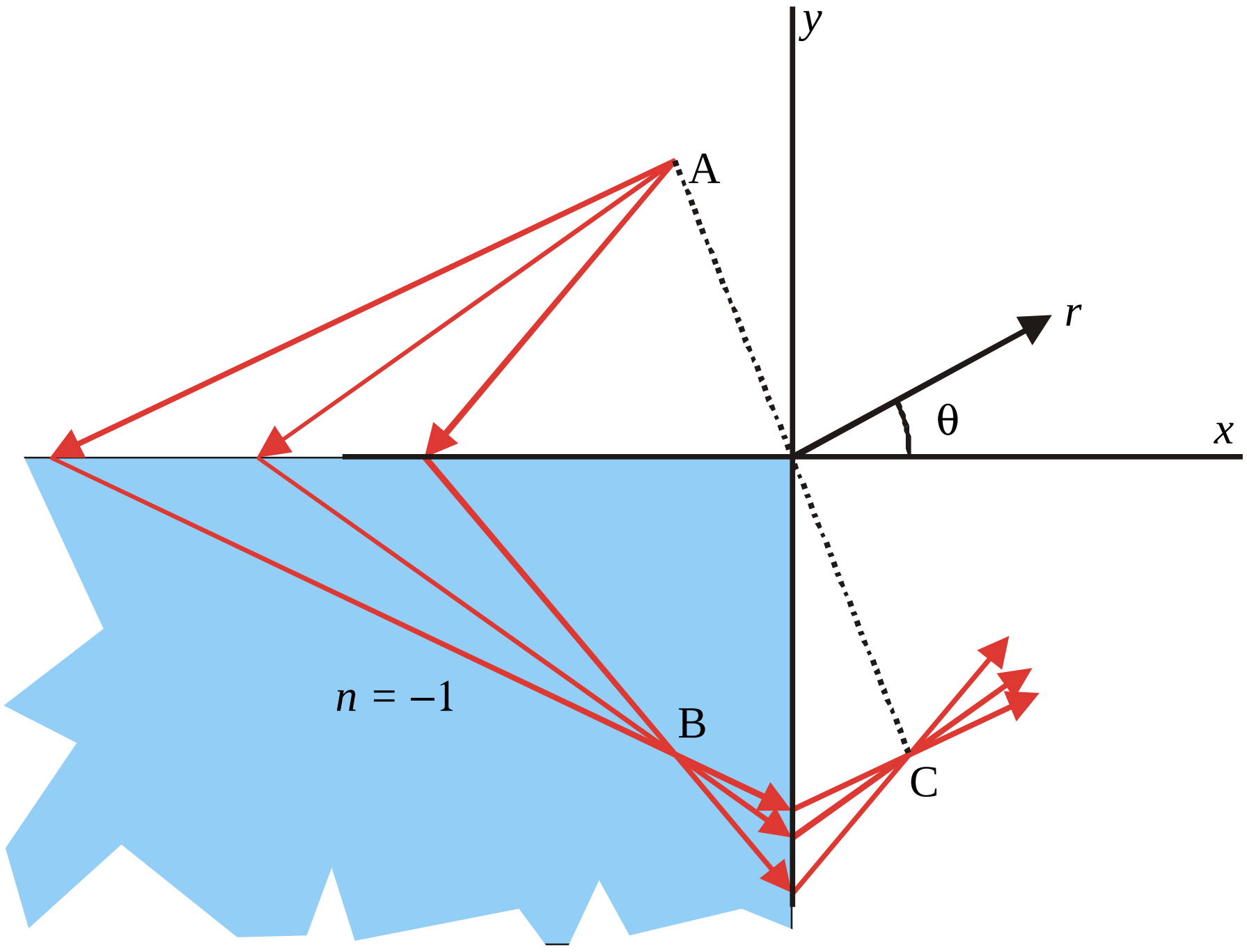

**Figure 9.** A corner reflector constructed from negatively refracting material, $n = -1$. Note that for each incident wave, the direction of the wave vector is reversed. This is not equivalent to time reversal because the rays do not return to the point of origin, but appear to radiate from point C, rotated by 180deg about the corner from the true origin at A. We show that this corner reflector is 'perfect' in the sense that the images at B and C include the correct near field components.

good analogy: a corner made of $\lim n \to -1$ material also reverses the *xy* components of the wave vectors. This wave vector reversal is quite distinct from time reversal as can be seen from the figure: the rays are not returned to their point of origin but have a different focus. In contrast to the corner mirror, this second focus is in free space external to the corner

The ray argument establishes the principles of the negative corner for propagating rays, but does the corner share the 'perfect' property of some other negatively refracting lenses? We use the powerful technique of coordinate transformation to prove that it does.

We start from the properties of the material in figure 9:

$$\begin{aligned} \varepsilon(\phi) = \mu(\phi) = +1, &\quad -\tfrac{\pi}{2} < \phi < \pi \\ \varepsilon(\phi) = \mu(\phi) = -1, &\quad -\pi < \phi < \tfrac{\pi}{2} \end{aligned} \tag{48}$$

then introduce a mapping of coordinates that takes the structure of figure 9 into the structure shown in figure 10.

$$x = r_0 \cos\phi\, e^{\ell/\ell_0}, \quad y = r_0 \sin\phi\, e^{\ell/\ell_0}, \quad z = Z \tag{49}$$

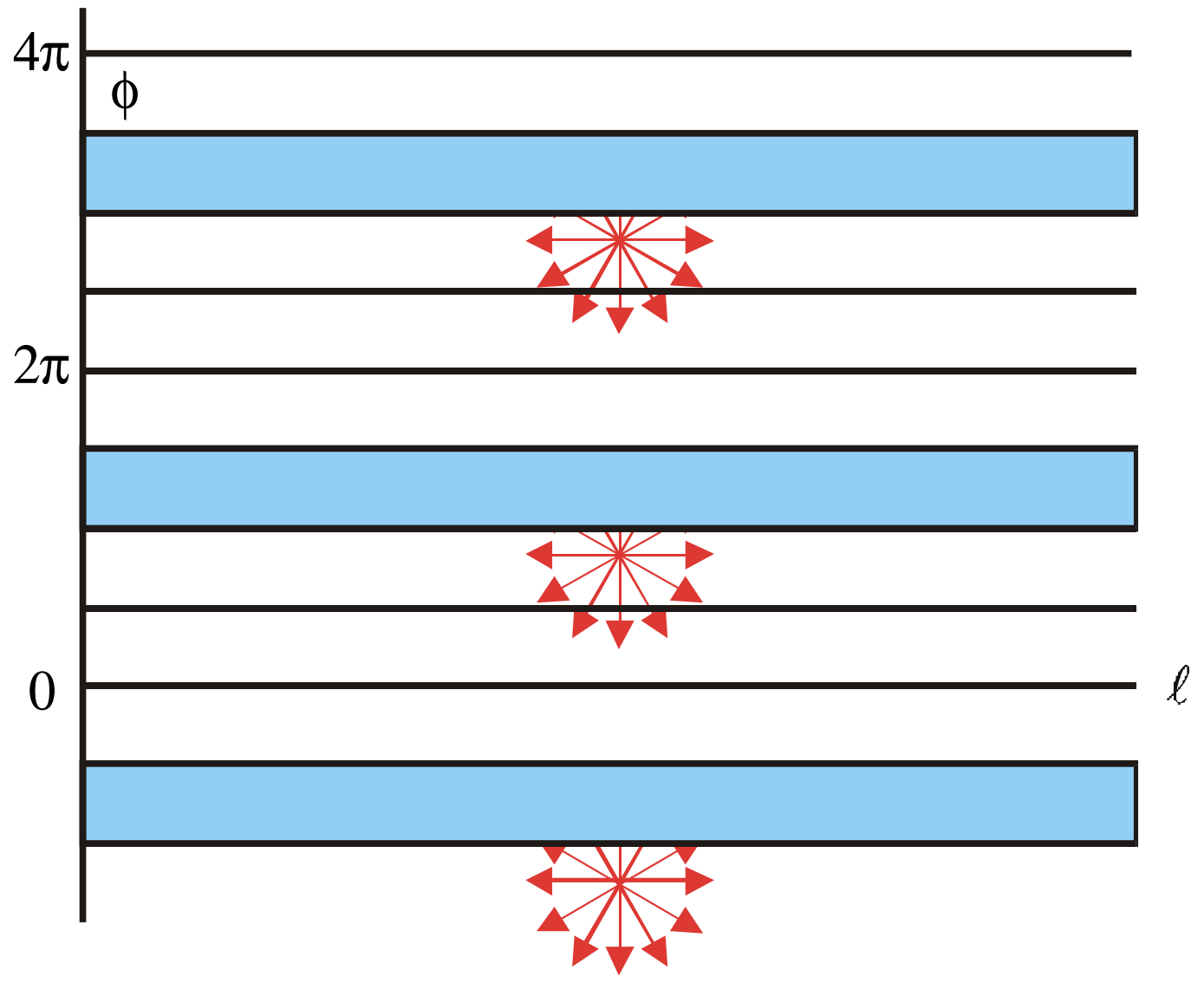

**Figure 10.** The previous figure can be mapped into a stack of slabs, where every fourth slab in the stack is complementary to the other 3. A point source is included in every fourth layer as shown so that the whole system is periodic in the $\phi$ direction.

In the new frame,

$$\tilde{\varepsilon}_i = \varepsilon_i \frac{Q_1 Q_2 Q_3}{Q_i^2}, \quad \tilde{\mu}_i = \mu_i \frac{Q_1 Q_2 Q_3}{Q_i^2} \tag{50}$$

where,

$$\begin{aligned} Q_\ell &= r_0/\ell_0 \sqrt{e^{2\ell/\ell_0}\cos^2\phi + e^{2\ell/\ell_0}\sin^2\phi} = r_0/\ell_0\, e^{\ell/\ell_0} \\ Q_\phi &= r_0 \sqrt{e^{2\ell/\ell_0}\sin^2\phi + e^{2\ell/\ell_0}\cos^2\phi} \quad = r_0 e^{\ell/\ell_0} \\ Q_Z &= 1 \\ Q_\ell Q_\phi Q_Z &= r_0^2/\ell_0\, e^{2\ell/\ell_0} \end{aligned} \tag{51}$$

and hence,

$$\begin{aligned} \tilde{\varepsilon}_\ell = \ell_0\, \varepsilon_\ell, \quad \tilde{\varepsilon}_\phi = \ell_0^{-1}\, \varepsilon_\phi, \quad \tilde{\varepsilon}_Z = r_0^2/\ell_0\, e^{2\ell/\ell_0}\, \varepsilon_z \\ \tilde{\mu}_\ell = \ell_0\, \mu_\ell, \quad \tilde{\mu}_\phi = \ell_0^{-1}\, \mu_\phi, \quad \tilde{\mu}_Z = r_0^2/\ell_0\, e^{2\ell/\ell_0}\, \mu_z \end{aligned} \tag{52}$$

Substituting from equation (48) and setting $\ell_0 = 1$ gives,

$$\begin{aligned} &\tilde{\varepsilon}_\ell = \tilde{\mu}_\ell = \tilde{\varepsilon}_\phi = \tilde{\mu}_\phi = +1;\ \tilde{\varepsilon}_Z = \tilde{\mu}_Z = +r_0^2 e^{2\ell} \quad -\tfrac{\pi}{2} < \phi < \pi \\ &\tilde{\varepsilon}_\ell = \tilde{\mu}_\ell = \tilde{\varepsilon}_\phi = \tilde{\mu}_\phi = -1;\ \tilde{\varepsilon}_Z = \tilde{\mu}_Z = -r_0^2 e^{2\ell} \quad -\pi < \phi < \tfrac{\pi}{2} \end{aligned} \tag{53}$$

Therefore the transformed medium satisfies the conditions of complementary media and the sources shown in figure 10 will each form two perfect images: one in the negative medium, and the other on the far side.

Hence it follows that the images in the corner reflector will also be perfect.

## 7. Two Negative Corners

Notomi [9] has observed that two negative corners will bend light around a loop and return rays to exactly the same state from which they started. The light forms a series of images as it circulates, and the fourth image has the identical location and phase as the source; see figure 11. Given an exact realisation of the $\lim n \to -1$ condition the light will circulate for ever. We can easily show that in addition to the propagating rays focussing, all the evanescent states also focus to give images in the ideal case of $\lim n \to -1$

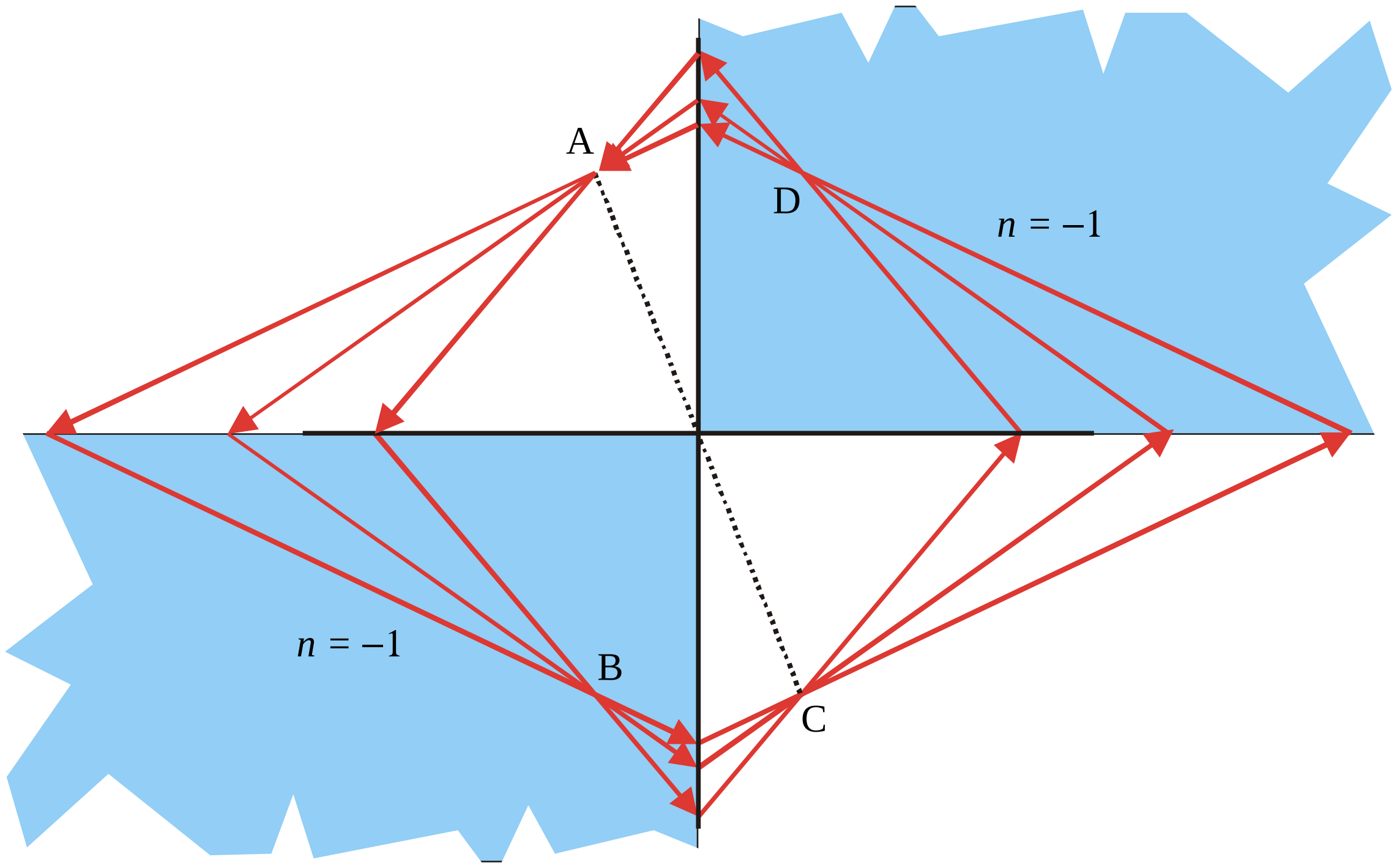

**Figure 11.** A double negative corner constructed from negatively refracting material, $n = -1$. Note that this configuration *cannot* be accomplished with mirrors. A sub set of the rays diverging from a source is eventually returned to the source and circulates around the system for ever. After three focussing events at B, C, D, the rays return to the source point, A, with the same phase with which they started out.

We construct our proof starting from figure 10 in the previous section where we considered the single corner. Using the same transformation double corner gives figure 12 below, the difference being that every second slab now contains negative material. Since this system consists of complementary slabs, it exhibits perfect lensing. Therefore we have demonstrated that the double negative corner produces perfect images. However the fields inside this structure are unusually singular and with time grow to infinity almost everywhere in the limit of zero absorption.

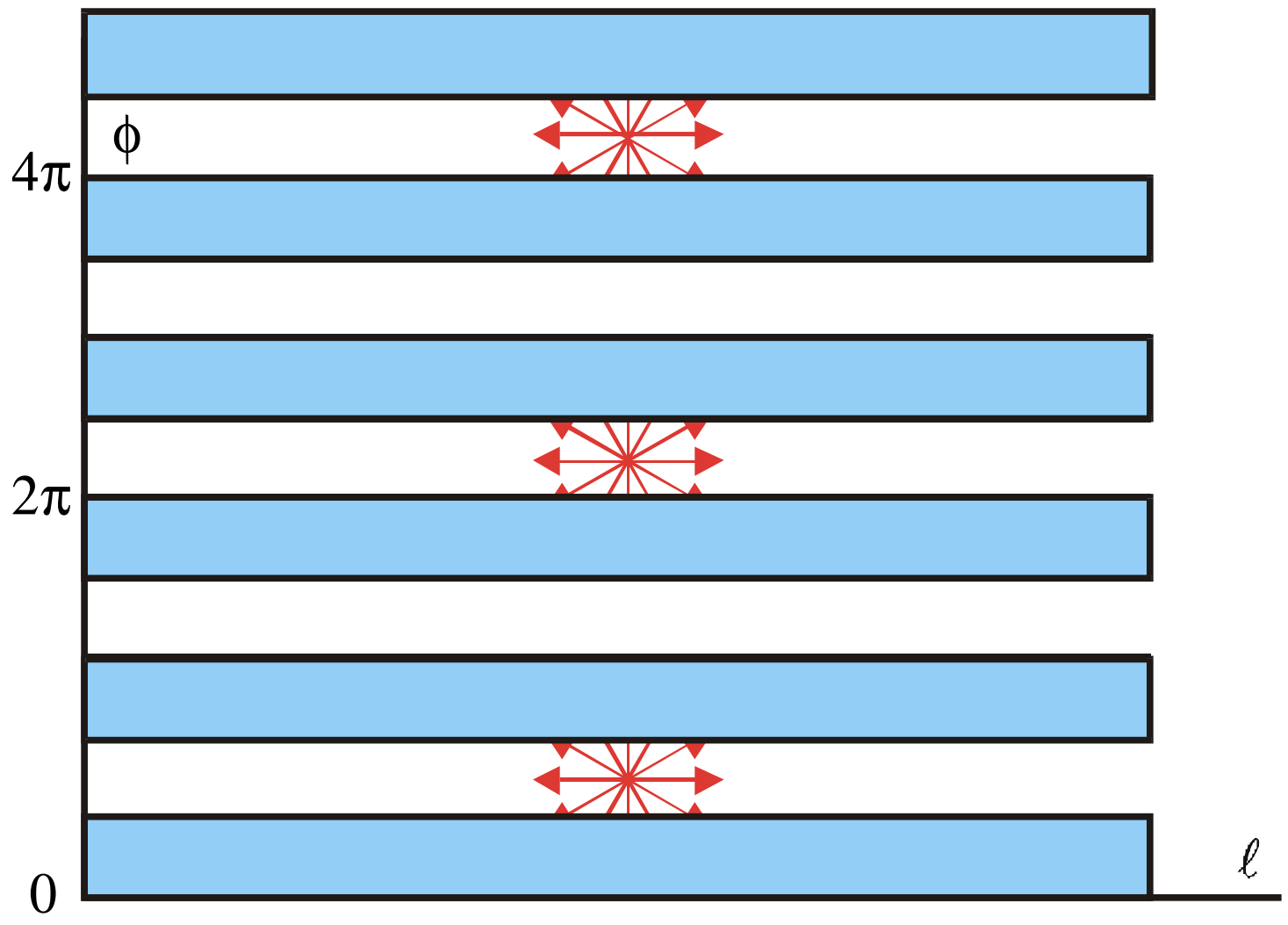

**Figure 12.** The previous figure can be mapped into a stack of slabs, where every succeeding slab in the stack is complementary to the previous one. A point source is included in every fourth layer so that the whole system is periodic in the *y* direction.

The ray diagram implies that there is a strict division between states which are trapped in an infinitely circulating loop, and states which head directly into free space. However this is only the case in the classical limit. A wave picture tells us that all states will have a finite chance of escaping from the corner.

Obviously the consecutive superimposed images lead to some unusual properties of the double negative corner. These can most simply be analysed in the short wavelength limit where the problem factorises into separate electro- and magneto-static problems. We also recognise that any system taking negative values of $\varepsilon$ or $\mu$ must also show dispersion with frequency. Typically we might have a plasma form for,

$$\varepsilon = 1 - \frac{\omega_p^2}{\omega(\omega + i\eta)} \tag{54}$$

and we are interested in the $\lim_{\eta \to \infty} \varepsilon$.

In the electrostatic limit, Apell et al [10] has proved a theorem for the surface plasmon modes [11]. If $\omega_{sp1}$ are the surface mode frequencies of our system, and $\omega_{sp2}$ are the modes of a second system in which the regions of plasma and dielectric are interchanged, then,

$$\omega_{sp1}^2 + \omega_{sp2}^2 = \omega_p^2 \tag{55}$$

In a class of systems where the exchange process maps the system into itself, this result defines the plasma modes. For example a half space filled with plasma and interfaced with vacuum maps into itself and therefore,

$$\omega_{sp1}^2 = \omega_{sp2}^2 \tag{56}$$

Our system also maps onto itself under the exchange and therefore,

$$\omega_{sp} = \omega_p \big/ \sqrt{2} \tag{57}$$

In other words in our system in the electrostatic limit all surface plasma modes are degenerate at $\omega_p \big/ \sqrt{2}$. At this frequency the density of states must be infinite. Also this is the frequency at which the condition $\varepsilon = -1$ is met.

In the limit we can easily write down the modes. The electrostatic potential of the modes has the form,

$$\begin{aligned} \psi &= \exp\left(+\alpha_x x - \alpha_y y + ik_z z\right), \quad x < 0, \quad y > 0 \\ &= \exp\left(-\alpha_x x - \alpha_y y + ik_z z\right), \quad x > 0, \quad y > 0 \\ &= \exp\left(-\alpha_x x + \alpha_y y + ik_z z\right), \quad x > 0, \quad y < 0 \\ &= \exp\left(+\alpha_x x + \alpha_y y + ik_z z\right), \quad x < 0, \quad y < 0 \end{aligned} \tag{58}$$

where,

$$\alpha_x^2 + \alpha_y^2 = k_z^2 \tag{59}$$

ensures that the modes obey Poisson's equation.

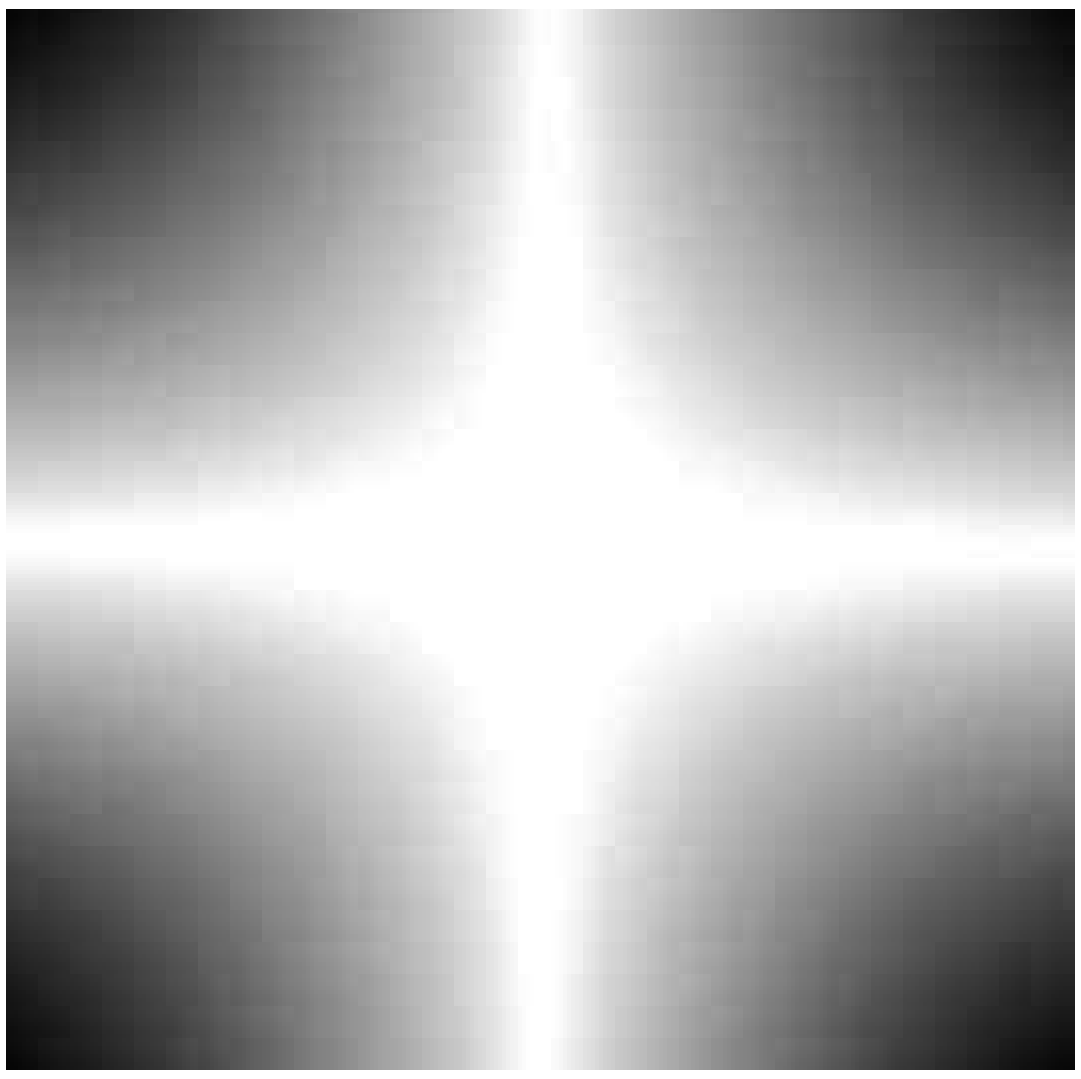

**Figure 13.** A sketch of one of the degenerate surface modes showing strong intensity at the interfaces, and especially in the corners.

As described the modes are not orthonormal, but an alternate orthonormal description is,

$$\begin{aligned} \psi &= \exp\left(+\alpha_x x - ik_y y + ik_z z\right), \quad x < 0, \quad y > 0 \\ &= \exp\left(-\alpha_x x - ik_y y + ik_z z\right), \quad x > 0, \quad y > 0 \\ &= \exp\left(-\alpha_x x + ik_y y + ik_z z\right), \quad x > 0, \quad y < 0 \\ &= \exp\left(+\alpha_x x + ik_y y + ik_z z\right), \quad x < 0, \quad y < 0 \end{aligned} \tag{60}$$

where,

$$\alpha_x^2 = k_y^2 + k_z^2 \tag{61}$$

Obviously these modes create a very high density of states near any of the surfaces, and especially in the corners of the system. This is the origin of the infinite induced response of the system at the surface plasmon frequency.

## 8. Conclusions

We have introduced the concept of ‘complementary media’ to describe the action of a perfect lens. Complementary media optically cancel one another at a specific frequency and as far as that frequency is concerned are invisible to the system. The concept is generalised from a planar slab of negatively refracting material to much more complex geometries related to one another by mirror symmetry. By introducing coordinate transformations the concept is extended to curved surfaces with a spheres given as a specific example. We discuss the corner reflector and show how its properties of reflection extend to evanescent as well as propagating waves. Two corner reflectors combine to make a most unusual cavity containing an infinity of degenerate resonances. We draw attention to the power of coordinate transformations as a tool for analysing negatively refracting and other optical materials.

## Acknowledgements

We acknowledge support from DoD/ONR MURI grant N00014-01-1-0803.

## Appendix

Starting from,

$$\nabla\times\left[\mu^{-1}(r)\nabla\times\mathbf{E}\right]=+i\omega\mu_0\nabla\times\mathbf{H}=+\omega^2c_0^{-2}\varepsilon(r)\mathbf{E} \tag{62}$$

we wish to derive an equation for the radial components of $\mathbf{E}$ and $\mathbf{H}$. These two components are sufficient to determine the entire field. Rearranging this equation,

$$-\frac{\mu'(r)}{\mu(r)}\hat{\mathbf{r}}\times\nabla\times\mathbf{E}+\nabla\times\nabla\times\mathbf{E}=+\omega^2c_0^{-2}\varepsilon(r)\mu(r)\mathbf{E} \tag{63}$$

We note that,

$$\nabla\times\mathbf{E}=\begin{bmatrix}\frac{1}{r\sin\theta}\left[\frac{\partial\left(\sin\theta E_\phi\right)}{\partial\theta}-\frac{\partial E_\theta}{\partial\phi}\right]\\ \frac{1}{r\sin\theta}\frac{\partial E_r}{\partial\phi}-\frac{1}{r}\frac{\partial\left(rE_\phi\right)}{\partial r}\\ \frac{1}{r}\left[\frac{\partial\left(rE_\theta\right)}{\partial r}-\frac{\partial E_r}{\partial\theta}\right]\end{bmatrix} \tag{64}$$

and,

$$\begin{aligned}&(\nabla\times\nabla\times\mathbf{E})_r\\
&=\frac{1}{r\sin\theta}\left[\frac{\partial}{\partial\theta}\left\{\sin\theta\frac{1}{r}\left[\frac{\partial\left(rE_\theta\right)}{\partial r}-\frac{\partial E_r}{\partial\theta}\right]\right\}-\frac{\partial}{\partial\phi}\left\{\frac{1}{r\sin\theta}\frac{\partial E_r}{\partial\phi}-\frac{1}{r}\frac{\partial\left(rE_\phi\right)}{\partial r}\right\}\right]\\
&=\frac{1}{r\sin\theta}\left[+\frac{\partial}{\partial\theta}\left\{\sin\theta\frac{1}{r}\frac{\partial\left(rE_\theta\right)}{\partial r}\right\}-\frac{1}{r}\frac{\partial}{\partial\theta}\sin\theta\frac{\partial E_r}{\partial\theta}-\frac{1}{r\sin\theta}\frac{\partial^2 E_r}{\partial^2\phi}+\frac{1}{r}\frac{\partial^2\left(rE_\phi\right)}{\partial\phi\partial r}\right]\\
&=+\frac{1}{r^2\sin\theta}\frac{\partial}{\partial\theta}\left\{\sin\theta\frac{\partial\left(rE_\theta\right)}{\partial r}\right\}+\frac{1}{r^2\sin\theta}\frac{\partial^2\left(rE_\phi\right)}{\partial\phi\partial r}-\frac{1}{r^2\sin\theta}\frac{\partial}{\partial\theta}\sin\theta\frac{\partial E_r}{\partial\theta}-\frac{1}{r^2\sin^2\theta}\frac{\partial^2 E_r}{\partial^2\phi}\\
&=-\frac{1}{r^3}\frac{\partial}{\partial r}\left(r^2\frac{\partial\left(rE_r\right)}{\partial r}\right)+\frac{1}{r}\nabla\cdot\frac{\partial\left(r\mathbf{E}\right)}{\partial r}-\nabla^2_{\theta\phi}E_r\end{aligned} \tag{65}$$

*Calculating* $\nabla\cdot\partial\mathbf{E}/\partial r$

Consider,

$$\nabla\cdot\mathbf{D}=\nabla\cdot\varepsilon(r)\mathbf{E}=\varepsilon'(r)\hat{\mathbf{r}}\cdot\mathbf{E}+\varepsilon(r)\nabla\cdot\mathbf{E}=0 \tag{66}$$

Explicitly,

$$\begin{aligned}\frac{\partial}{\partial r}\nabla\cdot\mathbf{E}&=\left[\frac{\partial}{\partial r}\nabla\right]\cdot\mathbf{E}+\nabla\cdot\left[\frac{\partial}{\partial r}\mathbf{E}\right]\\
&=\left[\frac{-2}{r^3}\frac{\partial}{\partial r}r^2E_r+\frac{2}{r^2}\frac{\partial}{\partial r}rE_r+\frac{-1}{r^2\sin\theta}\frac{\partial}{\partial\theta}\sin\theta E_\theta+\frac{-1}{r^2\sin\theta}\frac{\partial}{\partial\phi}E_\phi\right]+\nabla\cdot\left[\frac{\partial}{\partial r}\mathbf{E}\right]\\
&=\left[\frac{-1}{r^3}\frac{\partial}{\partial r}r^2E_r+\frac{2}{r^2}\frac{\partial}{\partial r}rE_r\right]+\nabla\cdot\left[\frac{\partial}{\partial r}\mathbf{E}\right]\\
&\quad+\left[\frac{-1}{r^3}\frac{\partial}{\partial r}r^2E_r+\frac{-1}{r^2\sin\theta}\frac{\partial}{\partial\theta}\sin\theta E_\theta+\frac{-1}{r^2\sin\theta}\frac{\partial}{\partial\phi}E_\phi\right]\\
&=\left[-\frac{2}{r^2}E_r-\frac{1}{r}\frac{\partial}{\partial r}E_r+\frac{2}{r^2}E_r+\frac{2}{r}\frac{\partial}{\partial r}E_r\right]+\nabla\cdot\left[\frac{\partial}{\partial r}\mathbf{E}\right]-\frac{1}{r}\nabla\cdot\mathbf{E}\\
&=\frac{1}{r}\frac{\partial}{\partial r}E_r+\nabla\cdot\frac{\partial}{\partial r}\mathbf{E}-\frac{1}{r}\nabla\cdot\mathbf{E}\end{aligned} \tag{67}$$

Substituting for,

$$\nabla \cdot \mathbf{E} = -\frac{\varepsilon'(r)}{\varepsilon(r)} E_r \tag{68}$$

gives,

$$-\frac{\partial}{\partial r}\left[\frac{\varepsilon'(r)}{\varepsilon(r)} E_r\right] = \frac{1}{r}\frac{\partial}{\partial r} E_r + \nabla \cdot \frac{\partial}{\partial r}\mathbf{E} + \frac{1}{r}\left[\frac{\varepsilon'(r)}{\varepsilon(r)} E_r\right] \tag{69}$$

Rearranging,

$$\nabla \cdot \frac{\partial}{\partial r}\mathbf{E} = -\frac{1}{r}\frac{\partial}{\partial r} E_r - \frac{\partial}{\partial r}\left[\frac{\varepsilon'(r)}{\varepsilon(r)} E_r\right] - \frac{1}{r}\left[\frac{\varepsilon'(r)}{\varepsilon(r)} E_r\right] \tag{70}$$

Recalling that,

$$(\nabla\times\nabla\times\mathbf{E})_r = -\frac{1}{r^3}\frac{\partial}{\partial r}\left(r^2 \frac{\partial(rE_r)}{\partial r}\right) + \frac{1}{r}\nabla \cdot \frac{\partial(r\mathbf{E})}{\partial r} - \nabla^2_{\theta\phi} E_r \tag{71}$$

we have,

$$\begin{aligned}\frac{1}{r}\nabla \cdot \frac{\partial(r\mathbf{E})}{\partial r} &= \frac{1}{r}\nabla \cdot \mathbf{E} + \frac{1}{r}\frac{\partial E_r}{\partial r} + \nabla \cdot \frac{\partial \mathbf{E}}{\partial r} \\ &= -\frac{1}{r}\frac{\varepsilon'(r)}{\varepsilon(r)} E_r + \frac{1}{r}\frac{\partial E_r}{\partial r} - \frac{1}{r}\frac{\partial}{\partial r} E_r - \frac{\partial}{\partial r}\left[\frac{\varepsilon'(r)}{\varepsilon(r)} E_r\right] - \frac{1}{r}\left[\frac{\varepsilon'(r)}{\varepsilon(r)} E_r\right] \\ &= -\frac{2}{r}\frac{\varepsilon'(r)}{\varepsilon(r)} E_r - \frac{\partial}{\partial r}\left[\frac{\varepsilon'(r)}{\varepsilon(r)} E_r\right]\end{aligned} \tag{72}$$

hence,

$$(\nabla\times\nabla\times\mathbf{E})_r = -\frac{1}{r^3}\frac{\partial}{\partial r}\left(r^2 \frac{\partial(rE_r)}{\partial r}\right) - \frac{2}{r}\frac{\varepsilon'(r)}{\varepsilon(r)} E_r - \frac{\partial}{\partial r}\left[\frac{\varepsilon'(r)}{\varepsilon(r)} E_r\right] - \nabla^2_{\theta\phi} E_r \tag{73}$$

Substituting into (**)

$$\begin{aligned}(\nabla\times\nabla\times\mathbf{E})_r &= -\frac{1}{r^3}\frac{\partial}{\partial r}\left(r^2 \frac{\partial(rE_r)}{\partial r}\right) - \frac{2}{r}\frac{\varepsilon'(r)}{\varepsilon(r)} E_r - \frac{\partial}{\partial r}\left[\frac{\varepsilon'(r)}{\varepsilon(r)} E_r\right] - \nabla^2_{\theta\phi} E_r \\ &= +\omega^2 c_0^{-2} \varepsilon(r)\mu(r) E_r\end{aligned} \tag{74}$$

which is the equation for $E_r$ which we sought. By symmetry there is a similar equation for $H_r$.